\documentclass[12pt]{iopart}

\usepackage{epsf}

\newcommand{\Sgmm}{\Sigma^0}
\newcommand{\Sgm}{$\Sgmm$\ }
\newcommand{\GSgm}{$\Sigma$\ }
\newcommand{\ASgmm}{\overline{\Sigma}^0}

\newcommand{\Gm}{$\gamma$}
\newcommand{\Lam}{$\Lambda$\ }

\newcommand{\Ratio}{$\Sgmm/\Lambda$\ }
\newcommand{\Piz}{$\pi^0$\ }
\newcommand{\ptarm}{$p_T^{AP}$}

\begin{document}

\title{Reconstructing \Sgm decays in STAR}

\author{G. Van Buren\dag\ for the STAR Collaboration\footnote{For
the full author list and acknowledgements, see Appendix ``Collaborations" in this volume.}
}
\address{\dag\ Dept. of Physics, Brookhaven National Laboratory, Upton, NY 11973-5000 USA}
\ead{gene@bnl.gov}

\begin{abstract}
Typical experimental measurements do not
separate the contributions of
\GSgm state decays from (anti)\Lam yields.
An ongoing analysis in STAR is
resolving the contribution from
excited \GSgm states~\cite{sevil}, but the primary contribution
comes from electromagnetic decays of the
(anti)\Sgm.

Photon conversions into $e^+e^-$
pairs in detector material have been used by STAR to
identify photons from \Piz decays~\cite{pi0}. A similar technique
has been used here to identify photons from (anti)\Sgm
decays in conjunction with STAR's excellent PID capabilities
for finding the associated (anti)\Lam daughters.
We report here on progress toward measuring the
(anti)\Sgm yields in various nuclear collisions at
RHIC.

\end{abstract}

\submitto{\JPG}
\pacs{25.75.-q, 25.75.Dw, 14.20.Jn}

\section{Introduction}

Decays of \GSgm states contribute to measurements of other baryons,
such as their inclusion in \Lam yields and spectra via
$\Sigma^0 \rightarrow \Lambda \gamma$
(with a branching ratio close to 100\%).  
Because the decay is electromagnetic, short decay lengths
make this measurement possible only by identification of
the partner \Gm~or its missing energy. Accurate particle
production model comparisons would benefit from understanding
such contributions.  A notable instance involves
one of the more prominent expectations of quark gluon plasma:
strangeness enhancement in the antihyperon channel measured via
$\overline{\Lambda}/\overline{p}$~\cite{enhance}.
Measurements often
entangle weak decay feeddown in both the numerator
and denominator,
like $\overline{\Sigma}^- \rightarrow \overline{p} \pi^0$.

Interpretation of \Lam spectra (such as $<$$p_T$$>$ versus mass
systematics)
may also suffer from not knowing \Sgm contributions.
Whether final state interactions of \GSgm baryons
in heavy ion collisions differ from \Lam is unknown as their cross sections
have been measured with very little overlap
(at $\sqrt{s} \approx 15$ GeV, where they
appear to agree with a value of approximately 33 mb~\cite{PDG1}).
Additionally, there exist data from only one experiment on how the
production of \Sgm and \Lam differ over phase space in nuclear collisions
($p$+$Be$ at $p_{lab} = 28.5$ GeV/$c$), where it appears
the ratio $\Lambda_{\Sigma^0 \rightarrow \Lambda\gamma}/\Lambda_{inclusive}$
is constant at $\sim$$\frac{1}{4}$.~\cite{pBe}.
Different feeddown from resonances into these states in larger
nuclear collisions may alter this. Any system size dependence
would impact observed trends like the linearity between \Lam and $h^-$ yields
seen in 130 GeV $Au$+$Au$ data at RHIC~\cite{Lam1}.

\subsection{\Ratio}

A better understanding of the physics of RHIC collisions will
come from measuring
relative yields of the \Sgm and \Lam.
It has been stated that isospin dictates that the ratio of the production
cross sections $\sigma(\Sgmm)/\sigma(\Lambda)$ should
be 1/3~\cite{cosy11}. Reasons for this are not obvious,
so we will examine other production arguments here.
To be clear, we will consider {\it final} yields to be after strong
(resonance) decays, but before electroweak decays.

\subsubsection{Models}

Using thermal model parameters fit from
central 200 GeV $Au$+$Au$ STAR data~\cite{Olga},
the THERMUS thermal model gives
a primordial ratio of 0.67, and a final
ratio of 0.36~\cite{THERMUS}.  The dependence of either of these values on the
parameters of the thermal model is very weak, but it is clear that the
resonances play a strong role. This is noteworthy because there are
indications that resonances are thermally under-populated (at least in the final state)
in the $Au$+$Au$ data, but reasonable in $p$+$p$~\cite{Olga,Markert}.

Simple counting rules in quark coalescence models, expected to be relevant only under
conditions of very dense matter with sub-hadronic degrees of freedom~\cite{qcoal,ALCOR},
lead to yield ratios of 1/1 for primordial production, but feeddown from
resonances into both species reduces this ratio to 1/5 if the
resonances are fully-populated~\cite{Levai}.
Here, the window is larger than for the thermal model.
If central $Au$+$Au$ collisions produce a \Ratio value near 0.2 or 1.0, 
this may serve as an indicator for quark coalescence.

Event generators  also have predictive power.
HIJING/$B\overline{B}$ has no final state rescattering to alter yields
of resonances~\cite{HIJING}, giving a value of 0.37 for the ratio in 200 GeV
$d$+$Au$ collisions, which agrees with the thermal model prediction for
a fully equilibrated system at RHIC including resonances.

\subsubsection{Past Results}

\begin{figure}
\begin{center}
\epsfxsize=6.5in
\epsfbox{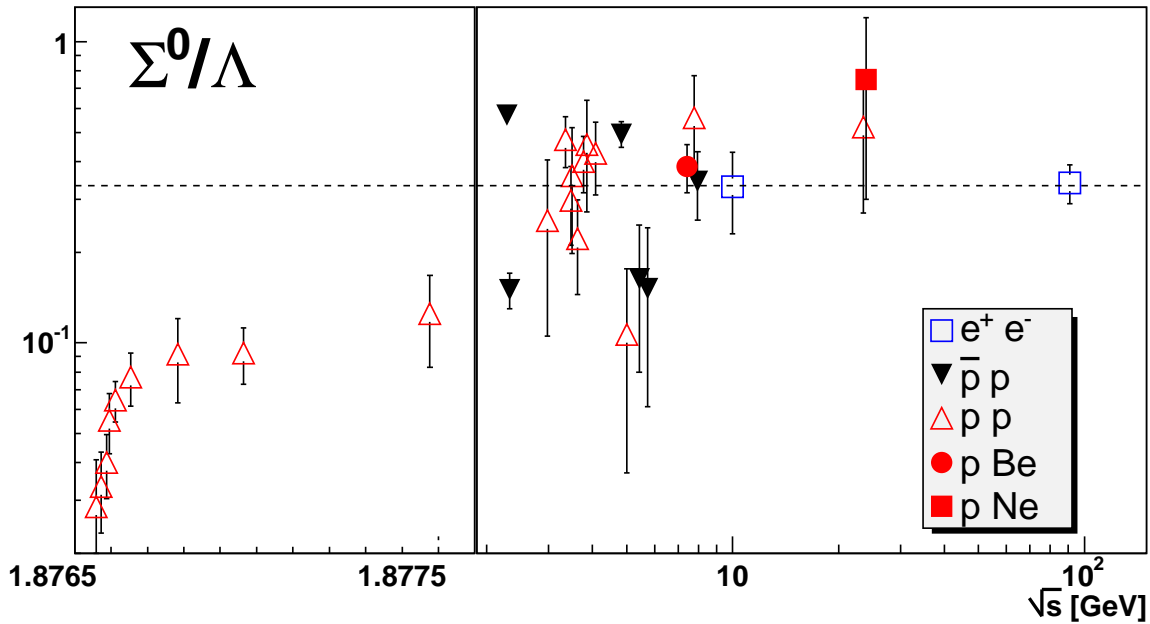}
\vspace{-0.8cm}
\end{center}
\caption{\label{fi:PastRes}\Ratio past results versus collision $\sqrt{s}$
($\sqrt{s_{NN}}$ for $p$+$A$)~\cite{pBe,past,pNe,Z0}. Meson-nucleon
reaction results are excluded for clarity, but exist only at intermediate energies
and lie in the same range.}
\vspace{-0.4cm}
\end{figure}

The low energy behavior of the \Ratio ratio shows a rise from threshold
for associated production of a \Sgm
($p$+$p \rightarrow p K^+ \Sgmm$),
as shown in Fig.~\ref{fi:PastRes}.
Measurements at intermediate energies show some spread, but generally
cluster near the isospin ratio of 1/3.
Only one data point exists from a colliding system
larger than $p$+$Be$, from $p$+$Ne$ at $\sqrt{s_{NN}} = 24$ GeV~\cite{pNe},
which has a
rather high value of $0.75 \pm 0.45$, but is consistent with 1/3 within errors.
This is insufficient to conclude whether larger colliding systems behave
differently. At higher energies, there exist only the measurements
of $Z^0$ decay products~\cite{Z0}, nothing from hadronic interactions.

\section{Experiment}

STAR has measured the  \Piz spectrum  in 130 GeV $Au$+$Au$ collisions
using  photon conversions ($\gamma \rightarrow e^+e^-$) in detector
material~\cite{pi0}.
While combinatorics leave the background under
the \Piz mass peak too large for identifying individual pions, the reconstructed
photons do have enough signal-to-noise to combine with
other potential decay products, as is true with \Lam reconstruction in STAR.
In the current \Sgm analysis, the topological similarities of photon
conversions to $V0$ (weak) decays was used to reconstruct the photons.
Two aspects uniquely identify
the conversions: daughter $d$E$/dx$ characteristics of electrons,
and very low values for the Armenteros-Podolanski decay variable \ptarm.
The electron $d$E$/dx$ band crosses other
$d$E$/dx$ bands in the low momentum range of interest,
so a cut on low \ptarm\ is more stringent. 

\begin{figure}
\begin{center}
\epsfxsize=4.5in
\epsfbox{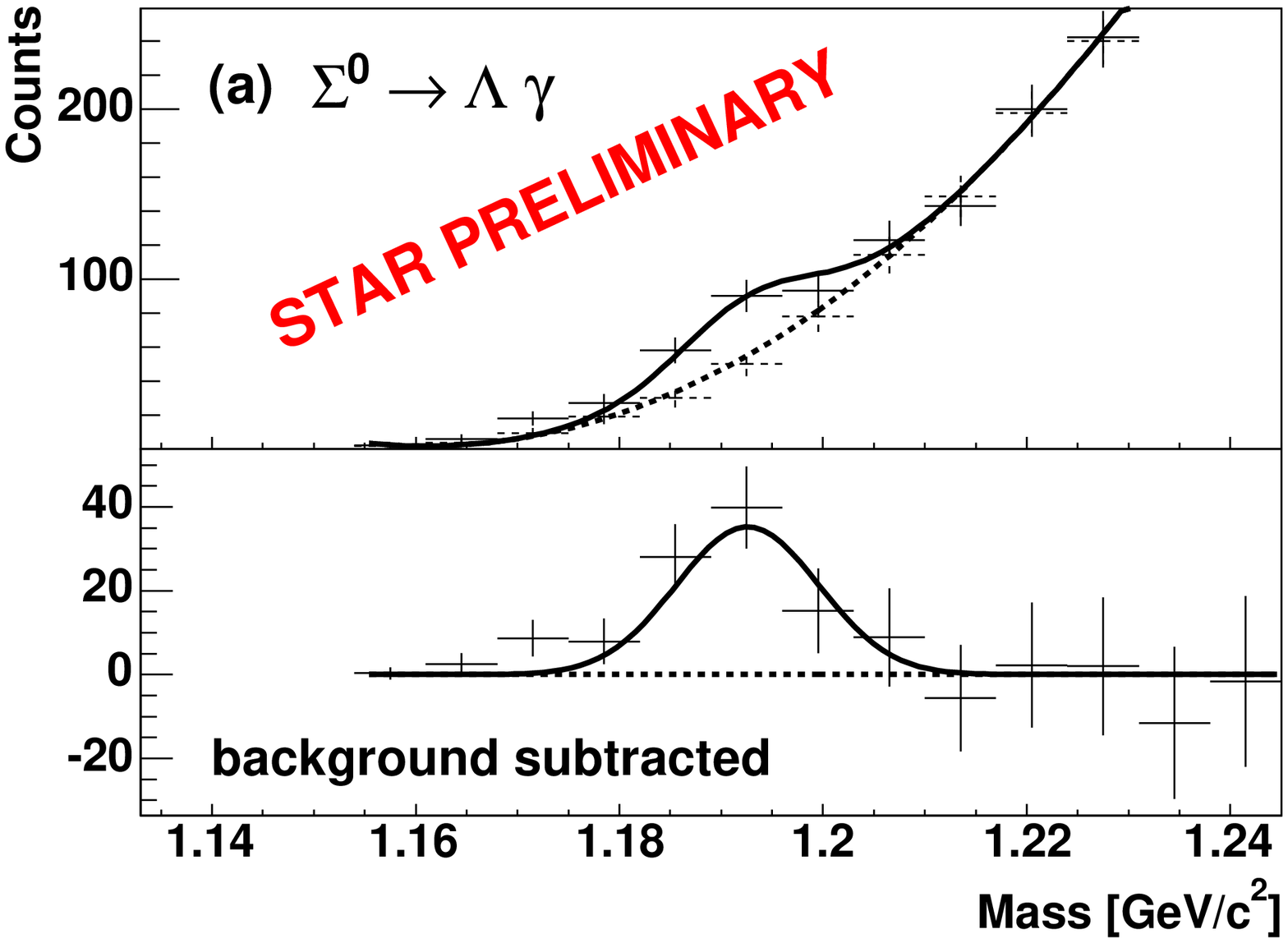}
\epsfxsize=4.5in
\epsfbox{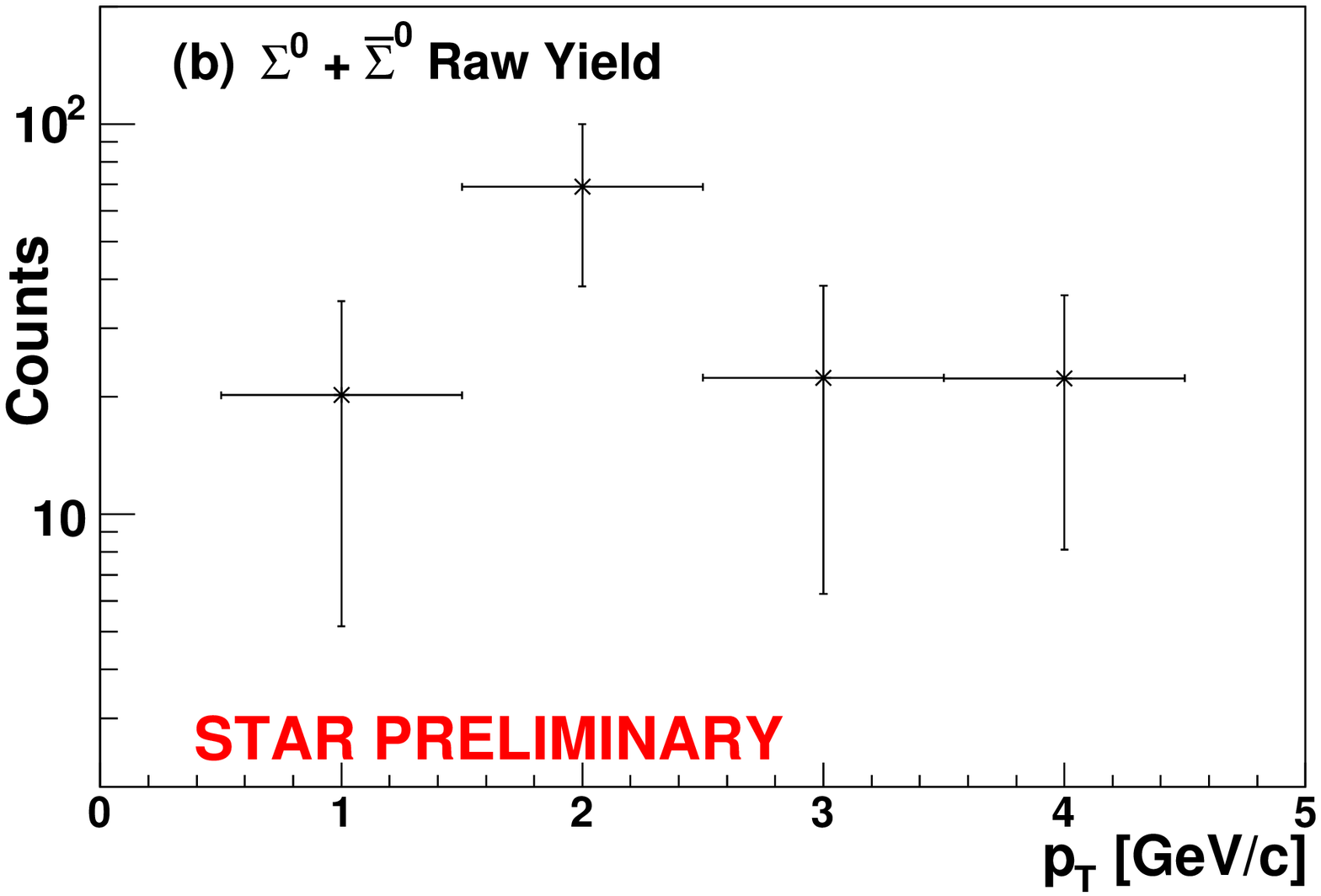}
\vspace{-1.1cm}
\end{center}
\caption{\label{fi:data} (a) Reconstucted \Sgm
invariant mass distribution (solid) and
rotated background approximation (dashed) with fits,
before and after background subtraction, and
(b) uncorrected counts of $\Sgmm$+$\ASgmm$
from mid-rapidity minimum bias 200 GeV $d$+$Au$.}
\vspace{-0.4cm}
\end{figure}

Only STAR data collected from minimum bias 200 GeV $d$+$Au$ collisions has
shown sufficiently low combinatoric backgrounds along with adequate statistics
to reconstruct a signal. Clear evidence of the \Sgm is shown from 14.7 million
minimum bias events in Fig.~\ref{fi:data}(a) for rapidities between $\pm 0.75$
over all $p_T$. A background shape has been determined by rotating \Sgm
daughter candidates about the beam axis. Systematic errors derived from
varying fit functions and ranges are dominated by statistical errors.
The resulting reconstructed counts (uncorrected for efficiency)
from four $p_T$ bins of $\Sgmm$+$\ASgmm$
are shown in Fig.~\ref{fi:data}(b). Using the uncorrected counts over
all $p_T$ we find $\ASgmm/\Sgmm = 0.6 \pm 0.3$. Future analyses
will benefit from using the more efficient \Gm-finder used in the
\Piz analysis, for which reconstruction efficiencies are better understood.

Weak decay contributions to the measurable \Sgm yield are minor, with
$\Xi^0$ and $\Xi^-$ each adding much less than 1\% of their own (relatively
small) yields. Strong decay contributions may be more significant, including
$12 \pm 2\%$ of the $\Sigma$(1385),
100\% and 40\% of $\Lambda$(1405) and $\Lambda$(1520) decays,
and uncertain fractions of heavier \GSgm and \Lam excited states.
Measurements of the $\Sigma$(1385) yield
in high energy nuclear collisions by STAR~\cite{sevil} will likely aid our understanding of
\Sgm sources.

\section{Summary}

We have observed a signal for (anti)\Sgm in STAR for
minimum bias 200 GeV $d$+$Au$ collisions at midrapidity,
and expect to improve it via a different \Gm-finder
which has shown higher efficiencies, such that
a yield can be measured. Finding yields in other colliding
systems may also be possible with further data acquired by STAR:
improved statistics of 200 GeV $Au$+$Au$ data from 2004 may overcome
the large combinatorial background in that data; 62 GeV $Au$+$Au$ data
taken in 2004 is expected to have reduced combinatorial background,
as should $Cu$+$Cu$ data to be taken in 2005; and future $p$+$p$ runs should provide
sufficient statistics to observe the (anti)\Sgm signals.

\end{document}